
\input phyzzx

\def\equation{$$}
\def\endequation{$$}

\def\KM{V.A. Kazakov and A.A. Migdal, Nucl. Phys. {\bf B397}  (214) 1993}

\def\orig{V.A. Kazakov and A.A. Migdal, Ref \KM; A.A. Migdal, Mod. Phys. Lett.
	{\bf A8} (359) 1993; A.A. Migdal, Mod Phys. Lett. {\bf A8}
	(153) 1993}

\def\Hch{Harish-Chandra, Amer. J. Math. {\bf 79} (87) 1957;
 	C.Itzykson and J.B. Zuber, J. Math. Phys. {\bf 21} (411) 1980;
 	M.L. Mehta, Comm. Math. Phys. {\bf 79} (327) 1981}

\def\elitzur{ S. Elitzur, Phys. Rev. {\bf D}12, 3978.}

\def\zn{I. Kogan, G. Semenoff, N. Weiss, Phys. Rev. Lett 69 (3435) 1992}

\def\fermion{A.A. Migdal, Mod. Phys. Lett. A8 (359) 1993}

\def\sublead{Phys.Lett. {\bf B302} (283) 1993. }

\def\gross{D.Gross, {\it Phys.Lett.} {\bf B293} (181) 1992}

\def\dms{DMS paper M.I. Dobroliubov, Yu. Makeenko, G.W. Semenoff,
	Mod. Phys. Lett. A8 P.2387 1993; Yu Makeenko, Mod. Phys. Lett {\bf A8}
	(209) 1993; Yu Makeenko, Mod. Phys. Lett. {\bf B422} (237) 1994}

\def\pensln{Yu. Makeenko, ``Critical Scaling and Continuum Limits
	in the D>1 Kazakov-Migdal Model'' hep-th 9408029; See also
	Ref. \usxyz}

\def\us{L.D. Paniak ``Singular Matrix Models and the Kazakov--Migdal
	Model ''  M.Sc. Thesis, University of British Columbia,
	(in preparation).  L.D. Paniak, N. Weiss, ``Kazakov--Migdal Model
	with Logarithmic Potential and the Double Penner Matrix Model''
	HEP-TH 9501037}

\def\lk{S. Chauduri, H. Dykstra, J. Lykken, Mod. Phys. Lett A V6 1665 1991}

\def\cd{{\cal D}}

\centerline{\fourteenrm \bf
 Status of the Kazakov--Migdal Model}
\vskip 1truein
\centerline{\twelverm  \bf Nathan Weiss}
\footnote{\dagger} {Talk presented at {\lq\lq Quarks '94\rq\rq}
May 1994, Vladimir, Russia
}
\centerline{\tenpoint \it Department of Physics, University of British
Columbia}
\centerline{\tenpoint \it Vancouver, B.C., V6T2A6, Canada}
\vskip .6truein
\centerline{\twelverm \bf ABSTRACT}
\tenrm
\baselineskip 12pt
\hsize 5.5truein
\hangindent .5truein
\parindent .75 truein

In this talk I discuss both the present status
and some recent work on the Kazakov--Migdal Model
which was originally proposed as a soluble, large $N$ realization of QCD.
After a brief description of the model and a discussion of its
solubility in the large $N$ limit I discuss several of the serious
problems with the model which lead to the conclusion that it
does {\it not} induce QCD. The model is nonetheless a very interesting
example of a Gauge Theory and it is related to some very interesting
Matrix Models. I then outline a technique \REF\dmsxyz{\dms}\refend
 which uses ``Loop Equations''
 for solving such models.
A  Penner--like model is then discussed with two logarithmic
singularities. This model is distinguished by the fact that it
is exactly and explicitly soluble in spite of the fact
that it is not Gaussian.  It is shown how to analyze this
model using both a technical approach and from a more physical
point of view.

\vskip .5truein
\twelverm
\baselineskip 14pt
\hsize 6truein

\section{Introduction}

The Kazakov--Migdal Model is an SU(N) Lattice Gauge model which
was proposed several years ago \REF\KMxyz{\KM}\refend
in the hope that it would be both soluble in the large $N$ limit and
that it would  give the correct long distance behavior of QCD
in the continuum limit. This model was thus also known as ``Induced
QCD''. Although it is now widely believed that this model does
$not$ induce QCD much has been learned from the study of this
model, both about ordinary and about gauged Matrix Models.
Work is still in progress on a class of ``soluble'', non--Gaussian
Kazakov--Migdal Models.

This talk will begin with a brief review of the Kazakov--Migdal Model
(the KM Model) and why it was originally though that it might induce
QCD. This will be followed by a discussion of the solubility
(i.e. tractability) of the model at large $N$.
Some of the main problems with the model will then be
described including a discussion of
the extra local $Z_N$ symmetry and the absense of critical behaviour
in the Gaussian Model. This will then explain why it is extremely
unlikely that this model could induce QCD.
This will be followed by an outline of a powerful method for solving
KM Models. I shall conclude this
talk with a discussion of a soluble, non--Gaussian, (``Penner'') model
which is explicitly soluble in the large N limit and which is the
focus of continued research.

\section{The Model}

The KM Model is a d--dimensional lattice Gauge Theory containing an SU(N)
Gauge Field $U_{xy}$ defined on the links of the lattice and a Hermetian
(adjoint) scalar ``Higgs'' Field $\Phi_x$ defined on the sites of the
lattice.  The model is defined in terms of the Euclidean Partition
Function
\equation
	Z~=~\int \prod_{links} \cd U_{xy} \prod_{sites} \cd \Phi_{x}
	{\rm exp} \left[ -S \right]
	\eqn{\Epi}
\endequation
with
\equation
	S=N\sum_{sites,x}{\rm tr}\left(V(\Phi_x)\right)
	-N\sum_{links,xy}{\rm tr}\left( \Phi_x U_{xy}\Phi_y
	U^{-1}_{xy}\right)
	\eqn{\Eaction}
\endequation
where $V(\Phi)$ is, at this stage, an unspecified potential for the field
$\Phi$ and where the second term is the standard gauge invariant kinetic
term for the adjoint field $\Phi$. Note that this action is simply
the lattice action for an adjoint scalar field {\bf except} that the
Wilson term which is the kinetic term for the Gauge Field is missing.
This Wilson term
\equation
	\sum_{plaquettes,1234}{\rm Tr}\left(U_1U_2U_3U_4\right) + h.c.
	\eqn{\Ewilson}
\endequation
is missing in order to assure the solubility of the model at large $N$.

The original hope was that the integration over $\Phi$ would
induce a kinetic term for the Gauge Field which would lead to
a theory which, in the continuum limit, would be QCD (or, more
precisely, Quantum Gluodynamics, since quarks have not been introduced
at this stage).  The arguments supporting this claim can be found in
the original papers \REF\origxyz{\orig}\refend.

\section{Large $N$ Solubility}

The missing Wilson term {\Ewilson} in Eq. {\Eaction}
allows for $exact$ integration over all the Gauge Fields $U_{xy}$ on
the lattice. To see this note that each term in the action contains
one and only one $U_{xy}$ and each such Gauge Field appears
in only one term in the action. The integral over $U$ can be done
by using the formula \REF\Hchxyz{\Hch}\refend
\equation
	\int  \cd U \exp(\sum_{i,j}\phi_i\psi_j \vert U_{ij}\vert^2)
 	= {\rm const} {{\det_{ij} e^{\phi_i\psi_j}}\over
	{\Delta(\phi)\Delta(\psi)}}
	\eqn{\Eiz}
\endequation
where
\equation
	\Delta(\phi)=\prod_{i<j}\left(\phi_i-\phi_j\right)
	\eqn{\Evander}
\endequation
is the Vandermonde determinant for $\phi$. This formula can be applied
to the integral in Eq. {\Epi} by diagonalizing the Hermetian Matrices
$\Phi_x$ and $\Phi_y$ which appear in Eq. {\Eaction}.

The main point is that the result of the integral depends only on the
eigenvalues of the $\Phi$'s. The integral over each matrix $\Phi$ (which
is an integral with the Hermetian measure) can now be written as
an integral over the eigenvalues $\phi_i$ of $\Phi$ since
\equation
	\int \cd \Phi \cdot\cdot\cdot ~~~~=~~~~\int \prod_i d\phi_i
	\Delta^2(\phi) \cdot\cdot\cdot
	\eqn{\Eevint}
\endequation
when the integrand depends only on the eigenvalues of $\Phi$.
It follows that the Partition Function, Eq. {\Epi}, is of the form
\equation
	Z~=~ \int \prod_{x,i} d\phi_{x,i}
	~\Delta^2(\phi_x)~ ~{\rm exp}
	\left[ - S_{\rm eff} \right]
	\eqn{\Eseff}
\endequation
where $S_{\rm eff}$ depends only on the eigenvalues of the $\Phi$'s.
In fact it contains one term which involves the $\phi_{x,i}$ at
neighbouring sites and a potential term involving only one site at a time.

In the limit of large $N$ the above integral can be done by finding
the {\bf Classical Minimum} of the action $S_{\rm eff}$. This is assumed to
occur when the eigenvalues of $\Phi_x$ are independent
of $x$. The basic problem is then to {\bf minimize} the effective action
\equation
	-{\rm log}\left\{ \left( {{{\rm det_{ij}}~{\rm e}^{N\phi_i\phi_j}}
	\over{\Delta^2(\phi)}}\right)^d \Delta^2(\phi)~{\rm e}
	^{-N{\rm Tr}V\left(\phi\right)}\right\}
	\eqn{\Eseffexplicit}
\endequation
with respect to all the $\phi_i$.
The result will be a specific set of eigenvalues $\phi_1 ... \phi_N$
which, in the limit of large $N$, is described by a density of
eigenvalues $\rho(\phi)$ (proportional to the number of eigenvalues
in the vicinity of $\phi$) which is normalized so that
$\int d\lambda \rho(\lambda)=1$.
This minimization can be carried out in practice for only a
very limited number of special choices for the potential $V$.
We shall discuss this further later in this talk.

\section {Problems with ``Induced QCD''}

Despite the solubility of the KM Model at large $N$, several serious
problems were recognized immediately with the hypothesis that the
KM Model induces QCD.
The first of these problems is the presence of an additional {\bf local}
$Z_N$ symmetry in the model\REF\znxyz{\zn}\refend.
 The absense of the Wilson term
allows us to multiply the Gauge Field $U_{xy}$ on ${\rm \underline {any}}$
link $(xy)$  by an element of the center of
the Gauge Group without affecting the action. More formally, the
transformation
\equation
	U_{xy} \rightarrow {\rm exp}\left({{2\pi ik}\over N}\right)
	U_{xy}
	\eqn{\Ezn}
\endequation
(which is also an element of SU(N) if $k$ is an integer) leaves
the action Eq. {\Eaction} invariant even if a different $k$ is
chosen for each link $(xy)$. (A Wilson term would ruin this
invariance.)

The main problem with this symmetry is that it implies the vanishing
of all Wilson Loops since
\equation
	<\prod_{loop} U> =  {\rm exp}\left({{2\pi ik}\over N}\right)
	<\prod_{loop} U>
	\eqn{\Ewloops}
\endequation
if we multiply any one of the links in the loop by
${\rm exp}\left({{2\pi ik}\over N}\right)$.
This forces all Wilson Loops to vanish.
Furthermore, since the symmetry is local, it cannot be broken
\REF\elitzurxyz{\elitzur}\refend. This is a disaster if the
model is to induce QCD since, in QCD, we expect an area law for
Wilson Loops.

Several proposals have been made to solve this problem, none of
which have been particularly successful. Attempts at solving this problem have
included the introduction of Fundamental Representation Fermions
\REF\fermionxyz{\fermion}\refend or a subleading (in $1/N$) Wilson
term \REF\subleadxyz{\sublead}\refend both of which break the
local $Z_N$ symmetry at the expense of the exact solubility of the model.
It has also been suggested that ``Filled Wilson Loops'' which
are invariant under the local $Z_N$ replace the
ordinary Wilson Loops in this model.

A second very serious problem with the idea of inducing QCD from the KM
Model was discovered when Gross \REF\grossxyz{\gross}\refend
solved the Gaussian KM Model, i.e. the model with
$V(\phi)=m^2\phi^2/2$, explicitly. The argument is as follows.
It turns out that the perturbative
mass of the ``Higgs'' scalar is $m^2-2d$ ($d$ is the dimension of the
spacetime.) The most convincing arguments that the KM Model induced
QCD were made for precisely this Gaussian
potential for which it was expected
that the model would have critical behaviour and a continuum
limit when $m^2 \rightarrow 2d$. The exact solution to the model
proved that no such critical behaviour was present and thus
the Gaussian model certainly did {\bf not} induce QCD.

This, by itself, does not rule out that some non-Gaussian model may
induce QCD but, in light of the fact that the main argument that the
model induced QCD did work for the Gaussian case and in light of
the $Z_N$ problem it seems very unlikely that any non-Gaussian
model would induce QCD. There are, in fact, several other problems
with the idea of Induced QCD which are discussed in the literature.

Despite the failure of the model to induce QCD, the KM Model is still
a very interesting Gauge Theory and it is very interesting to
find non-Gaussian models which can be solved exactly.
Fortunately there is a Penner-like model with a logarithmic
potential which can be solved explicitly (Ref. {\dmsxyz})
and which will be discussed below.

\section{ A Technique for Solving KM Models}

In this section I shall review briefly the method of Dobroliubov, Makeenko
and Semenoff (Ref. {\dmsxyz}) for finding the extremum of the
action {\Eseffexplicit}.  The basic idea is to begin with the
original action {\Eaction} and to consider the following two quantities:
\equation
	E(\lambda)=\left<{1\over N}{\rm Tr}
	\left({1\over{\lambda-\Phi_x}}\right)\right>
	= \int d\phi {\rho(\phi) \over {\lambda-\phi}}
	\eqn{\Eelambda}
\endequation
and
\equation
	G(\lambda,\nu)= \left<{1\over N}{\rm Tr}
	\left({1\over{\lambda-\Phi_x}}U_{xy}
	{1\over{\nu-\Phi_y}}U^{-1}_{xy}\right)\right>
	\eqn{\Egmunu}
\endequation
where the average is with respect to the original Path Integral and
where the expression of $E(\lambda)$ in terms of the density of eigenvalues
$\rho$ is valid in the large $N$ limit.

The main idea is to write a set of equations, similar to Schwinger--Dyson
Equations and often called ``Loop Equations'' for these quantities.
These equations are of the form
\equation
	\int \cd U \cd \Phi ~ {d\over {d\Phi_{x,ij}}}\left(
	\left[{1\over{\lambda-\Phi_{x}}}\right]
	_{ij}{\rm e}^{-S}\right)~=~0
	\eqn{\Esdelambda}
\endequation
and
\equation
	\int \cd U \cd \Phi ~ {d\over {d\Phi_{x,ij}}}\left(\left[
	{1\over{\lambda-\Phi_x}}U_{xy}
	{1\over{\nu-\Phi_y}}U^{-1}_{xy}\right]_{ij}
	{\rm e}^{-S}\right)~=~0
	\eqn{\Esdg}
\endequation
The first equation Eq. {\Esdelambda} will yield an equation for
$E(\lambda)$ and $G(\lambda,\nu)$ whereas the second equation will,
in general, involve quantities with more factors of $1/(\lambda-\phi)$.

In solving these equations one needs to make heavy use of the large
$N$ limit in which the integral is dominated entirely by a
single matrix $\Phi$ which minimizes the effective action
Eq. {\Eseffexplicit}. Even averages over the Gauge fields $U$ which
appear in Eq. {\Esdg} are performed in this background $\Phi$ field.
In fact it is useful to treat
\equation
	\int \cd U U_{(xy)}^{ij} \left(U_{(xy)}^{-1}\right)_{lm}{\rm e}^{-S}
	\eqn{\Euav}
\endequation
as an unknown whose only nonvanishing components are defined as
\equation
	C_{ij}={1\over Z}\int \cd U \vert U_{ij}\vert^2
	{\rm e}^{N~{\rm Tr}\Phi U \Phi U^{-1}}
	\eqn{\Ecij}
\endequation
where $\Phi$ is the minimum of the effective action which has yet
to be determined.
Obviously $C_{ij}$ depends on $\Phi$. It is also useful
to define a quantity
\equation
	\Lambda_i=C_{ij}\phi_j
	\eqn{\ELam}
\endequation
where $\phi_j$ are the eigenvalues of $\Phi$. Now the unknown but fixed
values of $\phi_i$ allow us to identify an integer with every eigenvalue
of $\Phi$. We can thus replace the quantity $\Lambda_i$ with a
quantity $\Lambda(\phi)$ which is defined via the correspondence
of $\phi$ with $i$.

The quantities defined above are very useful in solving the ``loop equations''
It turns out that there are several cases in which these
loop equations close and one can solve explicitly for $E(\lambda)$
and $G(\lambda,\nu)$. In these cases the density of eigenvalues
can be extracted from $E(\lambda)$ using Eq. {\Eelambda} which
implies that $E(\lambda)$ has a branch cut along the support of
$\rho$ whose discontinuity is proportional to $\rho$.
$C_{ij}$ can also be determined as the double discontinuity
across the cut of $G(\lambda,\nu)$ in $\lambda$ and in $\nu$.

The simplest case in which these equations close is the Gaussian
case in which $V(\phi)=m^2\phi^2/2$. This case is discussed
in detail in Ref. {\dmsxyz}. What is even more interesting is that
this system can be solved for a non-Gaussian, though singular potential
of the Penner \REF\penslnxyz{\pensln}\refend (logarithmic) type.
An outline of the solution follows.

Begin with the ``Loop Equations'' Eqs. {\Esdelambda} and {\Esdg}.
The first equation Eq. {\Esdelambda} leads to the equation
\equation
	E(\lambda)^2-\left<{{V^\prime(\phi)-2d\Lambda(\phi)}
	\over{\lambda-\phi}}\right>~=~0
	\eqn{\Ea}
\endequation
Here $V^{\prime} = dV/d\phi$
The second equation Eq. {\Esdg} can be simplified if we assume
a simple form for the quantity $V^{\prime}-2(d-1)\Lambda$.
The soluble ``Penner'' case consists of requiring this
quantity to have a simple pole at some point $\xi$ so that
\equation
	V^{\prime}(\phi)-2(d-1)\Lambda(\phi)
	= {q\over{\phi-\xi}}+B
	\eqn{\Eb}
\endequation
This ``ansatz'' allows a significant simplification of
the equations. Using the asymptotic conditions
\equation
	E(\lambda) \sim {1\over \lambda} + {{\bar\phi} \over {\lambda^2}}
	+ \cdot\cdot\cdot
	\eqn{\Ec}
\endequation
\equation
	G(\lambda,\nu) = {{E(\lambda)}\over \nu} + \cdot\cdot\cdot
	\eqn{\Ed}
\endequation
one eventually derives a quadratic equation for $E(\lambda)$ in terms
of a single, unknown quantity $\bar\phi$ which must be determined
self consistently. This quadratic equation for $E(\lambda)$ is simply
\equation
	E(\lambda)^2\left[(\lambda-B)(\lambda-\xi)\right]
	+E(\lambda)\left[ (\lambda-B)(\lambda-\xi)(\xi-B)
	+q(B-\xi)-(\lambda-\xi)\right]
\endequation
\equation
	+\left[ (\lambda+\bar\phi)(B-\xi)+(\xi^2-B^2)\right]=0
	\eqn{\Ee}
\endequation

Besides allowing us to solve for $E(\lambda)$ and thus for the density
of eigenvalues $\rho(\lambda)$ we can now find the potential which
leads to the ansatz {\Eb}. It is not difficult to show that if we were to
substitute the expression
\equation
	V^\prime(\lambda)-2d\Lambda(\lambda)=
	{q\over{\lambda-\xi}}+{{(1-q)}\over{\lambda-B}}
	+(\xi -B)
	\eqn{\Ef}
\endequation
into equation {\Ea} we would obtain
precisely the same quadratic equation (Eq. {\Ee}) for $E(\lambda)$
as Equation {\Ee}. It thus follows that
\equation
	\Lambda(\lambda)= {{(q-1)}\over{\lambda-B}}+\xi
	\eqn{\Eg}
\endequation
and thus the potential $V$ is given by
\equation
	V^\prime(\lambda)= {q\over{\lambda-\xi}}
	+{{(2d-1)(q-1)}\over{\lambda-B}}
	+B+(2d-1)\xi
	\eqn{\Eh}
\endequation

We are thus able to solve the KM Model for a potential with {\bf two}
logarithmic singularities plus a linear term \footnote{1}{There is actually
a subtle assumption being made here. It is not guaranteed that
$V$ will simply be the integral of $V^\prime$ unless $V$ is of the
special form ${\rm Tr}f(\phi)$ where $f$ is some ordinary function of
$\phi$. The point is that since we have made an ansatz for
$V^\prime - (2d-1)\Lambda$ rather than for $V$ we are not
guaranteed that $V$ will be of this simple form. }
when the various
coefficients are related as in Eq. {\Eh}.

\section {Relationship with an Ordinary Matrix Model}

We are now ready to see
that the solution to the above KM Model is closely related to
the solution of an ordinary Matrix model with the potential
$V^\prime-2d\Lambda$ given by Eq. {\Ef}. To see this consider an ordinary
matrix model defined as an integral over a single $N\times N$ Hermetian
matrix  $\Phi$
\equation
	Z~=~\int \cd \Phi {\rm exp}\left[-N{\rm Tr} W(\Phi)\right]
	\eqn{\Sa}
\endequation
This model can be solved by defining $E(\lambda)$ in analogy with
Eq. {\Eelambda} as
\equation
	E(\lambda)=\left<{1\over{\lambda-\Phi}}\right>
	\eqn{\Sb}
\endequation
and writing a ``loop equation'' in analogy with Eq. \Esdelambda which
leads to an equation identical to Eq. {\Ea}
\equation
	E(\lambda)^2-\left<{{W^\prime(\phi)}
	\over{\lambda-\phi}}\right>~=~0
	\eqn{\Sc}
\endequation
It follows that if, for $W^\prime$, we choose $V^\prime-2d\Lambda$
given by Eq. {\Ef} we arrive at precisely the same equation
for $E(\lambda)$ and thus the same distribution of eigenvalues
$\rho(\lambda)$.

We are thus led to consider an ordinary Matrix Model with
two logarithmic singularities so that $W$ is given by
\equation
	W^\prime(\lambda)={r_1\over{\lambda-\eta_1}}
	+{r_2\over{\lambda-\eta_2}}+C
	\eqn{\Oa}
\endequation
This model is related to the KM Model above provided (see Eq. {\Ef})
\equation
	r_1+r_2=1;~~~~C=\eta_2-\eta_1
	\eqn{\Ob}
\endequation

It is most interesting to consider the {\bf general} two pole Penner Model
above and to discuss the KM case as the special case when
Eq. {\Ob} is satisfied.
In the general  case Eq. {\Sc} can be written as:
\equation
	E(\lambda)^2-W^\prime(\lambda)E(\lambda)+\sum_{i=1}^2
	{{r_i E(\eta_i)}\over{\lambda-\eta_i}}=0
	\eqn{\Oc}
\endequation
Using the asymptotic expansion Eq. {\Ec} one can trade the two unknowns
$E(\eta_1)$ and $E(\eta_2)$ for $\bar\phi$ via the relations
\equation
	\sum_{i=1}^2 r_i E(\eta_i) =C
\endequation
\equation
	\sum_{i=1}^2 r_i E(\eta_i) \eta_i =
	r_1+r_2-1+C\bar\phi
	\eqn{\Od}
\endequation
The solution for $E(\lambda)$ is given by
\equation
	E(\lambda)={1\over 2}\left( W^\prime(\lambda) \pm
	\sqrt{W^\prime(\lambda)^2
	-4\sum_{i=1}^2
	{{r_i E(\eta_i)}\over{\lambda-\eta_i}}}\right)
	\eqn{\Oe}
\endequation

Now recall that $E(\lambda)=\int d\phi \rho(\phi)/(\lambda-\phi)$
has a branch cut precisely on the support of $\rho$. Furthermore on such a
branch cut of $E(\lambda)$
\equation
	E(\lambda+i\epsilon)-E(\lambda-i\epsilon)
	=2\pi i\rho(\lambda)
	\eqn{\Of}
\endequation
Thus the first step in finding the density of eigenvalues is to
find the location of the branch points of $E(\lambda)$ which
can be done by finding the places where the term inside the
square root vanishes. This is a quartic equation and thus, in general,
there will be two branch cuts. So the general solution
will be a two cut solution. Of course there will be cases when there
are no real solutions to the quartic equation and there will
be cases when we have a one cut solution.

The technical details of the various solutions for various values of the
parameters will be discussed in more detailed papers on this subject.
\REF\usxyz{\us}\refend
What I would like to discuss in this talk is a general physical
picture of what kind of solutions one should expect. Before doing
so, however, there is one further important technical point.
When any of the $r_i$ are positive the potential has an infinitely
attractive logarithmic well. In such cases it is known from the
single pole Penner Model \REF\lkxyz{\lk}\refend that the branch cuts
may have to circle the singularity. Mathematically this means that there
is no real solution for $\rho$ though physically it may be possible to
interpret this as a ``condensation of eigenvalues'' at the singularity.

\section{Physical Picture of Solutions}

It is instructive and not difficult to get a good physical picture
of the location of the eigenvalues of $\Phi$ for the solution of both
the simple Matrix Model and the KM Model. In particular it is not
difficult to decide whether solutions exist, whether they
are unique, where the eigenvalues are relative to the singularities
and whether there are one--cut or two--cut solutions.

We begin with the simple Matrix Model given by Equation {\Sa} which
can be written, using Eq. {\Eevint} as
\equation
	\int \prod_i d\phi_i \Delta^2(\phi)
	{\rm exp}\left[-N{\sum_i} W(\Phi_i)\right]
	\eqn{\Da}
\endequation
In large $N$, we must minimize the effective Action
\equation
	-\sum_{(i\ne j)}{\rm log}\left(\phi_i-\phi_j\right)^2
	+N\sum_i W(\phi_i)
	\eqn{\Db}
\endequation
Notice that the minimization of this Action corresponds to an
analogue classical mechanical problem in which a large number $N$
of particles are located at locations $\phi_1 ... \phi_N$ along
a line. They are each subjected to a central potential $W(\phi)$
and to a two--body repulsive potential ${\rm log}(\phi_i-\phi_j)^2$.
Thus, for example, if the potential has a single minimum at
some point $\phi_0$, one expects the eigenvalues to collect near
this minimum with some distribution which is controlled by the
logarithmic repulsion. If the potential has several minima then,
presumably, there are a whole class of {\bf classical} solutions
in which a varying proportion of particles are located in each of
the wells. In the Penner--like model in which there are at most two such
minima, these various cases can be distinguished by the value of
$\bar \phi$.

Let us now apply these ideas to the potential {\Oa}. The simplest case
occurs when both $r_1$ and $r_2$ are negative. Recall that Eq. {\Oa}
gives the $derivative$ of the potential so that in this case
the potential has two {\bf repulsive} singularities at
$\eta_1$ and $\eta_2$ and a linear term with either a positive or
a negative coefficient. In either case the potential has precisely
two minima. If, without loss of generality, we consider the case in
which $C$$>$$0$ then one minimum is between $\eta_1$ and $\eta_2$
and the other is at a point greater than $\eta_2$ (assuming $\eta_2>\eta_1$).
The potential is unbounded below for $\phi<\eta_1$. The fact that the potential
is unbounded does not affect the existence of solutions. In fact
{\bf classically} the analogue ``particles'' can exist in the two
minima without being significantly affected by what occurs for
$\phi < \eta_1$. Thus we expect a whole class of solutions which
can be parameterized by $\bar\phi$ which are, in general, two--cut
solutions, in which some of the eigenvalues live in a compact region
between $\eta_1$ and $\eta_2$
and the remainder live in another compact region
near the minimum beyond $\eta_2$.

The next case to consider is that in which both $r_1$ and
$r_2$ are positive. In this case the potential has two
{\bf attractive} singularities at $\eta_1$ and at $\eta_2$.
It has two maxima but no minima. In this case there are clearly
no nonsingular solutions to the analogue problem though, as discussed
previously, there may be some solutions in which the cuts of $E(\lambda)$
circle the singularities and which can be interpreted as
a condensation of eigenvalues. These turn out to be cases in which the
quartic form under the square root in Eq. {\Oe} has real solutions
but, in order to get both the asymptotic (large $\lambda$) behaviour
of $E(\lambda)$ and its behaviour near the singularities correctly
the cuts must circle one or more singularities.

The final case, in which the $r_i$ have opposite sign, is the most
interesting case for the KM Model. If, without loss of generality,
we assume that $\eta_1<\eta_2$ and $C>0$ then we identify two cases.
If $r_1<0$ and $r_2>0$ then the potential is repulsive at $r_1$ and
attractive at $r_2$. In this case we expect no real solutions, though
``condensation of eigenvalues'' may occur at $r_2$. In the case
$r_1>0$ and $r_2<0$ the potential is attractive at $r_1$ and
repulsive at $r_2$. Thus there is a minimum for $\phi>\eta_2$
and we should expect at least one real one--cut solution to this
problem for some specific value of $\bar\phi$.

To see that this last case is the one relevant to the KM Model recall
Equation {\Ob} which states that for the KM case $r_1+r_2=1$.
Now in order to get a real solution we need at least one of the
$r_i$ to be negative. In this case we see that the other one must be
positive so that the sum equals 1.  We see from the preceding
paragraph (and from the fact that $C=\eta_2-\eta_1>0$) that if
$r_2>0$ we expect no real solutions whereas if $r_2<0$ we
expect a single cut solution, where the cut is near the
minimum at $\phi>\eta_2$.

Let us now compare this with Eq. {\Ef}. We should identify
$\eta_2$ as $\eta$, $\eta_1$ as $B$ and $r_2$ as $q$.
Thus we expect a real single cut solution only if
$q<0$.  The form of the $KM$ potential can then be deduced
from Eq. {\Eh}. Note that the qualitative form of $V$ is
not uniquely determined. Even though $B-\xi<0$, the sign
of $B+(2d-2)\xi$ depends on the precise values of $B$ and $\xi$.
What is however uniquely determined is the sign of the singularities
(provided $d\ge 1$). If $q<0$ then $(2d-1)(q-1)<0$ and both
singularities {\bf in $V$ as opposed to $W$} are repulsive. The
integral over the Gauge Fields force a renormalization of the
analogue potential from $V$ to the effective Matrix Model form
$W$.

In the previous paragraphs we have shown how to estimate the
approximate location of the eigenvalues for an ordinary
Matrix Model or for a KM Model
 given the relationship between a KM Model
and an ordinary Matrix Model.
It is, however possible to get some information from the KM Potential
directly. To see this suppose we have a KM Model with an arbitrary potential
$V$. Our job is to find the extrema of the Effective Postential
given by Eq. {\Eseffexplicit}. This can be rewritten in the following form"
\equation
	S_{\rm eff}= N \sum_i \left( V(\phi_i)-d~\phi^2\right)
	+(d-1) \sum_{i\ne j} {\rm log}\left(\phi_i-\phi_j)\right)^2
	-d~{\rm log}\left( {\rm det}_{ij}{\rm e}^{N(\phi_i-\phi_j)^2/2
	}\right)
	\eqn{\De}
\endequation
Using the fact that the Integral {\Eiz} is finite and nonzero
when $\phi=\psi$ and when any two eigenvalues are equal,
it is easy to see that minimization of Eq. {\De} corresponds to
the following analogue mechanical problem. We imagine again $N$ particles
living on a line with a central potential which is
$V-d~\phi^2$ (not simply $V$), with an interparticle potential which
is no longer 2--body but which has the following property:
Whenever any two eigenvalues approach each other there is a repulsion
(the Action diverges) and whenever any two eigenvalues are separated
from each other by a large distance, there is an attraction which
tries to bring them closer together. This is quite different
from the physical picture which emerges from the simple Matrix Model.
\footnote \dagger {Unfortunately the compatibility between this picture and
the one which emerges from writing an effective one Matrix Model
and using the Physical Picture in that case, is not at all clear.
In fact, as mentioned briefly earlier in this article, there is some
possibility that the solution which corresponds to the one Matrix Model
is $not$ the KM Model with the simple potential ${\rm Tr V}$ given as
the integral of $V^\prime$ but one with a more complicated
matrix structure. These issues will be discussed in
a more detailed publication.}

\section{Summary}

At this point there is no evidence that the Kazakov--Migdal
Model provides a soluble, large $N$ realization of QCD and such
a hypothesis has many serious difficulties. We have, however,
learned much from the study of the KM Model both about Matrix Models
and about this new, interesting class of Gauge Theories.

\refout

\end